\pdfoutput=1
\documentclass[prl,aps,reprint,showpacs,showkeys,floats,floatfix,superscriptaddress, citeautoscript]{revtex4-1}

\usepackage{natbib}
\setcitestyle{super}
\AtBeginDocument{\nocite{achemso-control}}

\usepackage{graphicx}
\usepackage{dcolumn}
\usepackage[colorlinks,allcolors=black,citecolor=blue,urlcolor=blue]{hyperref}
\usepackage[dvipsnames]{xcolor}

\begin{document}

\title{Random sampling versus active learning algorithms for machine learning potentials of quantum liquid water}

\author{Nore Stolte}
\email{nore.stolte@theochem.rub.de}
\affiliation{Lehrstuhl f{\"u}r Theoretische Chemie, Ruhr-Universit{\"a}t Bochum, 44780 Bochum, Germany}
\author{J{\'a}nos Daru}
\affiliation{Lehrstuhl f{\"u}r Theoretische Chemie, Ruhr-Universit{\"a}t Bochum, 44780 Bochum, Germany}
\affiliation{Department of Organic Chemistry, E{\"o}tv{\"o}s Lor{\'a}nd University, 1117 Budapest, Hungary}
\author{Harald Forbert}
\affiliation{Center for Solvation Science ZEMOS, Ruhr-Universit{\"a}t Bochum, 44780 Bochum, Germany}
\author{Dominik Marx}
\affiliation{Lehrstuhl f{\"u}r Theoretische Chemie, Ruhr-Universit{\"a}t Bochum, 44780 Bochum, Germany}
\author{J{\"o}rg Behler}
\affiliation{Lehrstuhl f{\"u}r Theoretische Chemie II, Ruhr-Universit{\"a}t Bochum, 44780 Bochum, Germany}
\affiliation{Research Center Chemical Sciences and Sustainability, Research Alliance Ruhr, 44780 Bochum, Germany}

\date{\today}

\begin{abstract}
Training accurate machine learning potentials requires electronic structure data comprehensively covering the configurational space of the system of interest.
As the construction of this data is computationally demanding, many schemes for identifying the most important structures have been proposed.
Here, we compare the performance of high-dimensional neural network potentials (HDNNPs) for quantum liquid water at ambient conditions trained to data sets constructed using random sampling as well as various flavors of active learning based on query by committee.
Contrary to the common understanding of active learning,
we find that 
for a given data set size,
random sampling leads to smaller test errors for structures not included in the training process. 
In our analysis we show that this can be related to small energy offsets caused by a bias in structures added in active learning, which can be overcome by using instead energy correlations as an error measure that is invariant to such shifts.
Still, all HDNNPs yield very similar and accurate structural properties of quantum liquid water, 
which demonstrates the robustness of the training procedure with respect to the training set construction algorithm even when trained to as few as 200 structures.
However, we find that for active learning based on preliminary potentials, a
reasonable initial data set is important to avoid an unnecessary extension of the covered configuration space to less relevant regions.

\end{abstract}

\maketitle

\section*{Introduction}

In recent years, machine learning potentials (MLPs) have been frequently adopted to represent the potential energy surfaces of a wide range of molecules and materials \cite{P4885, Deringer2019Machine, Noe2020Machine, Behler2021Four, Deringer2021Gaussian, Friederich2021Machine, Unke2021Machine, Wen2022Deep, Kocer2022Neural}.
They are typically constructed to replace high-level electronic structure calculations,
as they are inexpensive to compute without significant loss of accuracy.
Thus, using MLPs, larger systems, greater timescales~\cite{lin2024machine}, and higher levels of theory~\cite{Daru2022Coupled,Chen2023Data-Efficient} are accessible in computational studies of atomistic systems than would be feasible with explicit electronic structure calculations.

The construction of an MLP requires a training set of atomic configurations labeled with their energies and/or forces at the target level of theory.
Beyond the choice of a machine learning method and its associated \mbox{(hyper-)parameters}, the performance of MLPs is determined by the quality of this training set. 
This is because, typically, MLPs perform well for structures which are close to those contained in the training set, but are not guaranteed to give equally good results for structures that are markedly different.
On the other hand, often the construction of the training set is the  computationally most intensive step in the development of an MLP, since demanding electronic structure calculations are required to label the training configurations.
For instance, an MLP trained on configurations of hexagonal ice may not do well for configurations of liquid water, although MLPs can have some degree of transferability.
For these reasons, a balanced data set has to be found containing a sufficient number of structures to provide the necessary information about the PES, while an excessively large number of structures soon becomes unaffordable.

Active learning
\cite{P5900}
is nowadays routinely used in the construction of training sets for MLPs
\cite{miksch2021strategies}
in order to address this issue. 
In an active learning procedure, the training set is iteratively extended to include more and more 
structures based on some predefined procedure until the quality of the potential converges.
Apart from this main application, active learning has also been used to decrease the training set size, thereby reducing
the training effort and possibly
the number of expensive electronic structure calculations required \cite{Raff2005AbIntio}.
Further, active learning can be used to explore reaction pathways \cite{Yang2022Metadynamics, Brezina2023Reducing, Zhang2024Modeling}, 
or to extend the applicability of the MLPs to a larger range of thermodynamic conditions \cite{Kapil2022First-Principles}.

When constructing a training set for an MLP,  first the source of training data has to be defined.
Active learning can be used to select atomic configurations from an existing database, consisting of e.g., a long \textit{ab initio} molecular dynamics simulation.
In that case, active learning is used to select the configurations that are necessary in order to represent the potential energy surface.
Where no 
database of reasonable (e.g., \textit{ab initio}) atomic configurations is available,
active learning procedures may rely on intermediate versions of the MLPs to sample new configurations, e.g., through equilibrium~\cite{Raff2005AbIntio, Malshe2007Theoretical, Artrith2012High-dimensional, Morawietz2012Neural, Sosso2012Neural, Morawietz2013Density-Functional, Morawietz2016How, Deringer2017Machine, Schran2020Automated, Schran2020Committee, Daru2022Coupled, Kapil2022First-Principles, Tang2023Machine, ONeill2024Pair}
or biased~\cite{Eshet2010Abinitio, David2024ArcaNN} molecular dynamics simulations, normal mode sampling~\cite{Smith2018Less}, or more sophisticated algorithms~\cite{Zhang2019Active, Guo2023ChecMatE}.
In addition to molecular dynamics simulations biased along one or more collective variables,
uncertainty-biased simulations with intermediate MLPs can expedite exploration of configurations that are not well-described in the training set~\cite{VanderOord2023Hyperactive, Zaverkin2024Uncertainty}.

Given an available pool of atomic configurations,
active learning can select structures for the training set in several ways.
Uncertainty predictions of an intermediate MLP can guide the selection, for example based on query by committee~\cite{Artrith2012High-dimensional, Smith2018Less, Zhang2019Active, Schran2020Committee, Smith2020ANI, VanderOord2023Hyperactive, Butler2024Machine}, Monte Carlo dropout~\cite{Thaler2024Active}, variance from Gaussian process regression~\cite{Zhai2020Active}, Bayesian inference~\cite{Jinnouchi2019Machine}, or other uncertainty measures~\cite{Zaverkin2021Exploration, Briganti2023Efficient, Zhu2023Fast, Zaverkin2024Uncertainty}.
Another approach is to select structures based on distances in feature space~\cite{Ludwig2007Abinitio, Dral2017Structure, Browning2017Genetic, Imbalzano2018Automatic, Huang2020Quantum, Sivaraman2020Machine, Lysogorskiy2023Active, Zhang2024Modeling, Zhang2024Active, Li2024Local}, distances in latent space~\cite{Janet2019Quantitative}, or a combination of both~\cite{Raff2005AbIntio, Malshe2007Theoretical}.
Combinations of feature space distance and predicted uncertainty have also been used to construct balanced training sets~\cite{Daru2022Coupled, Zaverkin2022Exploring, Tang2023Machine, Jung2024Active, Dubois2024Atomistic}.

Since MLP performance is dependent on the relationship between the structures in the training set and the relevant configurational space,
an understanding of the variation in performance of different active learning schemes is needed.
However,
comparisons of different approaches are rare and intermediate results with different data set sizes along active learning iterations are hardly reported~\cite{Tokita2023Tutorial}.
Nevertheless, a few recent studies have indicated that some active learning schemes may not outperform random sampling of the training set.
Active learning based on model uncertainty alone led to larger force errors 
than random training set selection for the aspirin molecule \cite{Zaverkin2022Exploring}, and alloys \cite{Gubaev2023Performance} 
for Gaussian moment neural network potentials.
One confounding factor is that for active learning schemes relying on an estimate of the MLP uncertainty, it is required that the actual MLP error and the MLP uncertainty are correlated, but that correlation is not necessarily strong, and the uncertainty may underestimate the actual error distribution \cite{Peterson2017Addressing, Kahle2022Quality, Zhu2023Fast, Tan2023Single, Zaverkin2024Uncertainty}.

\begin{figure}
\includegraphics{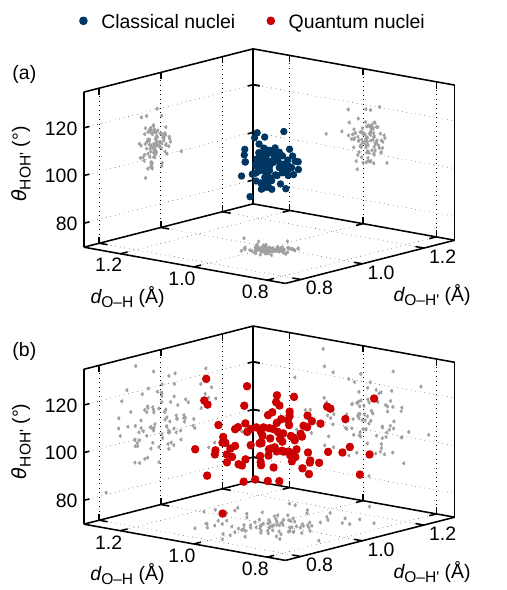}
\caption{
Intramolecular H$_2$O configurations from simulations with classical and quantum nuclei.
Here,
100 water molecules were randomly selected from classical (a) and path integral (b) constant-temperature constant-volume molecular dynamics simulations at 300 K and with experimental density at 1 atm, using the \mbox{RPBE-D3} density functional.
Points are placed in the scatter plot according to the two covalent bond lengths ($d_{\mathrm{O-H}}$ and $d_{\mathrm{O-H'}}$) and the bond angle $\theta_{\mathrm{HOH'}}$ in each molecule, and the projections of the points on each of the three planes are shown in gray. 
}
\label{configurations}
\end{figure}

In order to achieve a better understanding of the efficacy of active learning, in this work we have studied the construction of MLPs for bulk liquid water.
Due to its high relevance for many fields of science, liquid water is often used as a system for benchmarking new computational methods, including machine learning methods \cite{Omranpour2024Perspective}, and in fact
active learning has been employed to construct MLPs for water using various machine learning approaches \cite{Morawietz2016How, Morawietz2018Interplay, Schran2020Committee, Yao2020Temperature, Xu2020Isotope, Gartner2020Signatures, Sommers2020Raman, Schran2021Machine, Zhang2021Modeling, Zhang2021Phase,
Yue2021When, Zhang2022Deep, Daru2022Coupled, Malosso2022Viscosity, Huo2023Microscopic, Chen2023Data-Efficient, Zhai2023Short-blanket, Xu2023Accurate, Muniz2023Neural, CalegariAndrade2023Probing, Gomes-Filho2023Size, Lin2023Temperature, Maxson2024Transferable, MonterodeHijes2024Comparing, Ibrahim2024Efficient}.
In water, there is a complex interplay of very different  interactions: covalent bonds, hydrogen bonds, and long-range dispersion and electrostatic interactions.
Therefore, a high-quality potential energy surface for liquid water must describe well the different interaction strengths acting over various length scales. 
Furthermore, nuclear quantum effects play an important role in liquid water~\cite{Paesani2009Properties, Marx2010Aqueous, Ceriotti2016Nuclear},
and the balance between inter- and intra-molecular interaction strengths must be appropriately considered to obtain qualitatively correct nuclear quantum effects in path integral 
simulations~\cite{Habershon2009Competing, Daru2022Coupled}.
Particularly challenging for the training of MLPs for path integral molecular dynamics (PIMD) simulations of water is that the configuration space of water with quantum nuclei is substantially larger than that of water with classical nuclei, as is shown in Fig.~\ref{configurations} for the example of the intramolecular structure of 
water monomers in liquid water. 
At ambient conditions, with classical nuclei, the HOH angles are found in the range 90--120$^{\circ}$ and the bond lengths 0.9--1.1 {\AA}, while in PIMD simulations bond angles range from 80 to 130$^{\circ}$, and bond lengths from 0.8 to 1.3 {\AA}. 
It is therefore crucial that configurations of quantum water are included in the training sets for MLPs that are intended for use in PIMD simulations to avoid inaccuracies due to 
extrapolation \cite{Daru2022Coupled}.
As the use of efficient MLPs has made PIMD simulations of large systems and long time scales now much more accessible, it can be anticipated that nuclear quantum effects will be more routinely included in simulation studies as MLPs become more widely adopted.
A profound understanding of how to train MLPs suitable for PIMD simulations is thus of vital importance.
For these reasons, quantum liquid water presents a prototypical yet challenging system for the investigation of active learning for MLPs.

In this work we have trained high-dimensional neural network potentials (HDNNPs) \cite{Behler2007Generalized, Behler2021Four} with four distinct schemes to construct training sets for quantum liquid water 
deliberately restricted to 
ambient conditions, comparing active learning to random selection of training data.
Our target electronic structure level of theory is RPBE-D3 density functional theory \cite{Hammer1999Improved, Grimme2010Consistent}, which closely reproduces the experimental structure of bulk liquid water \cite{Forster-Tonigold2014Dispersion,Morawietz2016How}.
We assess how training sets constructed using different methods affect the HDNNP quality for structural properties, relative to liquid water from path integral \textit{ab initio} molecular dynamics (PI-AIMD).
These training and benchmarking efforts rely on a PI-AIMD simulation longer than 300 ps, and a grand total of more than 120 ns of PIMD simulations with HDNNPs.
Furthermore, we focus the training on bulk liquid water at one state point, 1 atm and 300 K, and evaluate the obtained HDNNPs against the reference PI-AIMD trajectory at the same conditions.
HDNNPs as well as other types of MLPs that are transferable across pressure/temperature conditions can and have been constructed, but this would introduce further degrees of freedom to the training set construction algorithms that would  obfuscate their impact on the performance of the potentials.
The illustrative examples of active learning schemes investigated here
cover active learning given a pre-existing \textit{ab initio} trajectory, active learning \emph{without} a pre-existing trajectory, as well as naive random sampling,
and we analyze in detail the structures contained in the training sets in each case and the consequences for the performance of the potentials.
Our results provide insights into the merits of active learning for MLPs, in particular for, but not limited to, aqueous solutions.

\begin{figure*}[ht]
\includegraphics[width=0.85\textwidth]{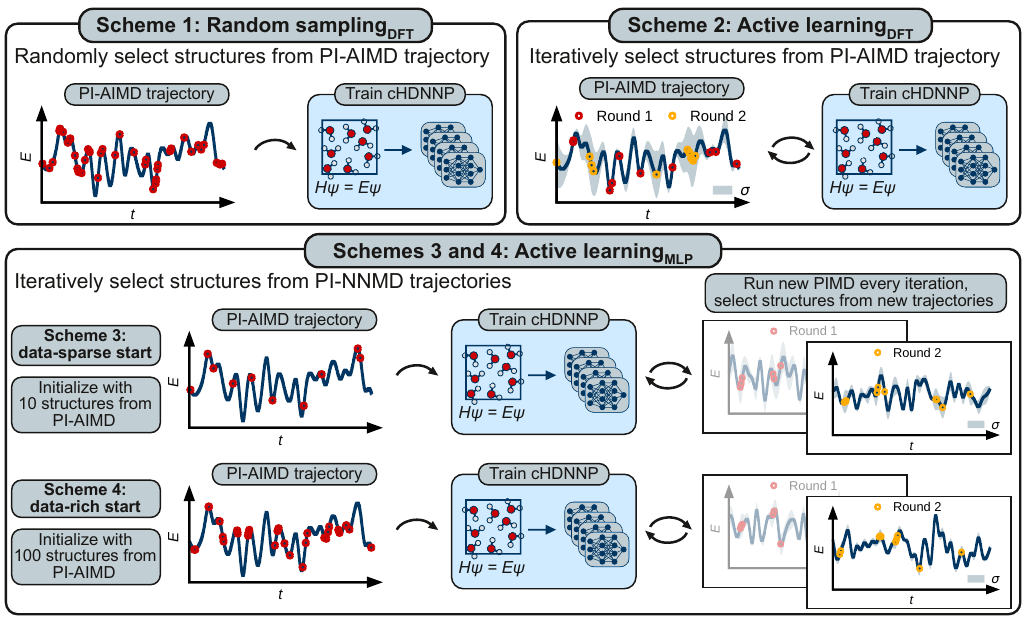}
\caption{
Overview of the four different schemes investigated in this work to construct training sets for cHDNNPs.
In scheme 1, data points are selected from a precomputed PI-AIMD trajectory by random sampling (RS$_{\mathrm{DFT}}$).
In scheme 2, data is selected from the same PI-AIMD trajectory, but starting from a first preliminary cHDNNP trained using a few points only, the number of data points is iteratively increased selecting further points with large committee errors from this DFT pool of structures.
by active learning (AL$_{\mathrm{DFT}}$).
In schemes 3 and 4, the first cHDNNP is constructed from a small (10 structures) or large (100 structures) initial data set, respectively.
New trajectories are then generated by cHDNNP-driven path integral molecular dynamics, and further batches of training structures are selected based on their committee errors to construct improved potentials iteratively.
Schemes 3 and 4 are labelled data-sparse active learning from MLP structures (AL$_{\mathrm{MLP}}$, data-sparse) and data-rich active learning from MLP structures (AL$_{\mathrm{MLP}}$, data-rich), respectively.
}
\label{schemes}
\end{figure*}

\section*{Methods}

\subsection*{\textit{Ab initio} molecular dynamics simulations}
We used CP2K \cite{Hutter2014CP2K, CP2K} with its Quickstep 
\cite{Vandevondele2005Quickstep,Kuhne2020CP2K} 
and path integral \cite{Brieuc2020Converged}
modules
to perform path integral \textit{ab initio} molecular dynamics (PI-AIMD) simulations \cite{MarxHutter2009}.
We performed 347.65 ps of constant-volume path integral molecular dynamics simulations of a periodic box containing 128 H$_2$O molecules, using 8 Trotter replica, referred to as beads, to discretize the path integral.
The temperature was maintained at 300 K with the path integral quantum thermal bath (PIQTB) thermostat \cite{Brieuc2016Quantum, Schran2018Converged}, and the density was fixed to its experimental value at 1 atm by setting the simulation cell size to 15.6627~{\AA}~\cite{Wagner2000IAPWS}.
Forces were evaluated using the RBPE exchange-correlation density functional \cite{Hammer1999Improved} with D3 dispersion \cite{Grimme2010Consistent} (using two-body terms and zero damping).
The density functional theory (DFT) calculations used dual-space norm-conserving Gaussian-type pseudopotentials \cite{Goedecker1996Separable} together with a Gaussian/plane wave (GPW) basis set for the valence electrons \cite{Lippert1997Hybrid}, using the triple-zeta TZV2P basis \cite{VandeVondele2007Gaussian} and a plane wave cutoff of 500~Ry. 
The time step in simulations was 0.25~fs.
The first 12.5~ps were considered equilibration and discarded, leaving a 335.15~ps trajectory.

The path integral RPBE-D3 trajectory, which we will refer to as the PI-AIMD trajectory, served two purposes. 
Firstly, radial distribution functions (RDFs) were computed for this trajectory, and the performance of HDNNPs in simulations was assessed by comparing HDNNP RDFs to those from this well-converged reference trajectory.
Secondly, the PI-AIMD trajectory served as the pool from which training structures were selected.
To this end, the equilibrated part of the trajectory was separated into two parts.
From the first 20 ps, structures were selected every 30~fs, each time randomly drawing one of the eight beads.
These 667 structures form a holdout test set.
No structures from the first 20~ps of the PI-AIMD trajectory were used in training.
From the remaining part of the trajectory, periodic frames were again drawn every 30~fs, now including all beads.
These frames formed the pool from which structures were selected during training.
In the rest of this text, to refer to the configurations of all 128 H$_2$O from the 8 beads in a single time step, we use the term ``frame"; 
when we refer to the configuration of a single bead only, containing all 128 H$_2$O molecules, we use the term ``structure".

In all learning schemes investigated in this work, the forces and energies in 
structures selected to be included in training sets were re-computed with the same settings as used in the RPBE-D3 PI-AIMD trajectory, except that the plane wave cutoff was increased to 1500~Ry. 
This minimizes noise in the training set, so that HDNNPs are not limited in their accuracy due to noise~\cite{Heid2023Characterizing}.
For a fair assessment of the HDNNPs, the energies and forces of the 667 frames in the test set 
were also recomputed with the 1500~Ry 
cutoff.

We performed path integral simulations with the HDNNPs as well, 
using the \mbox{RubNNet4MD} interface \cite{Schran2018High, Schran2020Automated, Brieuc2020Converged} of CP2K explained below, 
to assess the HDNNPs trained in this work, and as part of active learning schemes to increase the training set size. 
These simulations kept the same settings as the PI-AIMD simulation, with 128 H$_2$O molecules and 8 beads, at experimental density with constant volume, thermostatted to 300~K.
We refer to these as PI-NNMD simulations.

\subsection*{High-dimensional neural network potentials}
The methodology of high-dimensional neural network potentials has been described in great detail in the literature, and we refer the interested reader to several reviews on the topic
\cite{P4885, Behler2021Four, Tokita2023Tutorial}.
In this work, we have trained second-generation HDNNPs using the \mbox{RubNNet4MD} code~\cite{RubNNet4MD}.
The HDNNPs used two hidden layers with 30 nodes in each layer, with a hyperbolic tangent activation function.
The local environment of each atom is described with atom-centered symmetry functions (ACSFs) \cite{Behler2011Atom}.
The functional forms and parameters of all symmetry functions are given in the Supporting Information (SI).

During training, HDNNP weights were updated using the element-decoupled Kalman filter \cite{Blank1994Adaptive, Gastegger2015High-Dimensional}.
Both energies and forces were used during training; force updates were given a weight of 10, relative to energy updates.
However, to speed up training procedures, only 10 \% of 
the force components were randomly chosen during each 
epoch of neural network weight updating.

For each training procedure, we construct not a single HDNNP, but instead a committee \cite{Schran2020Committee}, consisting of 8 underlying HDNNPs with the same symmetry functions and architecture trained on the same reference data.
For a given training set, a 9:1 train:validation split was made at random for each committee member, and the weights of each committee member were initialized according to the Nguyen-Widrow scheme \cite{Nguyen1990Improving} with different random seeds.
Throughout the text, we refer to a committee of 8 HDNNPs as a single ``committee HDNNP", cHDNNP for short.

\section*{Results and Discussion}

\subsection*{Random versus active learning}

\begin{figure}
\includegraphics{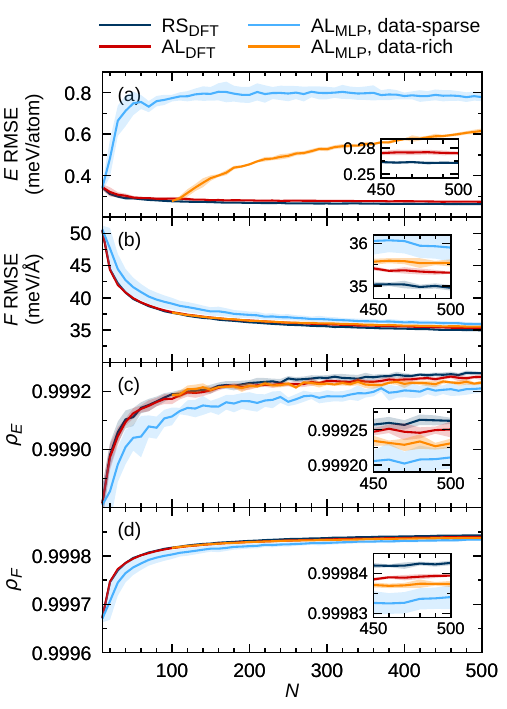}
\caption{
Quality of energy and force predictions of cHDNNPs on the test set of bulk liquid water structures, as a function of the number of 128-molecule structures $N$ in the training set.
Energy (a) and force (b) root mean squared errors (RMSEs), and energy (c) and force (d) Pearson correlation coefficients ($\rho$) between cHDNNP and RPBE-D3 data.
Solid lines show average
RMSEs (a, b) and correlations (c, d) of the 
cHDNNPs from the independent training procedures performed in each of the four active learning schemes.
Shaded areas indicate the standard deviation across those training procedures.
}
\label{correlation_rmse}
\end{figure}

During training, in
each active learning step
10 new 128-H$_2$O structures were added to the training set.
We first compared two selection schemes for these 
structures,
using random sampling and active learning (Fig.~\ref{schemes}).
In both these schemes, new structures were selected from the pool of training structures from the RPBE-D3 PI-AIMD simulation.
In the random sampling scheme (RS$_{\mathrm{DFT}}$), 
structures
were simply added at random to the cHDNNP training set,
although new 
structures
were only accepted into the training set if no other path integral beads from the same time frame were previously included,
since the configurations of different beads from the same time step in a path integral simulation are highly correlated. 
In the active learning scheme (AL$_{\mathrm{DFT}}$), the force committee disagreement was evaluated for all atoms of one bead per time frame of the pool of $\sim$10000 frames, according to
\begin{equation}
\sigma_{F_{\alpha}} = \left[ \frac{1}{n} \sum_{i=1}^n | \vec{F}_{\alpha,i} - \langle \vec{F}_{\alpha} \rangle |^2  \right]^{\frac{1}{2}},
\label{comdis}
\end{equation}
where $\alpha$ is the atom index, $n$ is the number of committee members, equal to 8 in this study, $\vec{F}_{\alpha,i}$ is the force predicted by the $i$th committee member on atom $\alpha$, and 
\begin{equation}
\langle \vec{F}_{\alpha} \rangle = \frac{1}{n} \sum_{i=1}^n \vec{F}_{\alpha,i},
\label{comforce}
\end{equation}
i.e., the 
mean force prediction, which is the committee force prediction.
For each time frame from the PI-AIMD simulation, a randomly selected path integral bead was used for evaluating the committee disagreement, again to avoid highly correlated structures to be added to the training set.
The maximum force committee disagreement among all atoms in a structure, $\max_{\alpha}\left(\sigma_{F_{\alpha}}\right)$, was determined, 
and the 10 structures that included the largest maximum force committee disagreements were added to the cHDNNP training set, before the cHDNNP was re-trained.

AL$_{\mathrm{DFT}}$ schemes were initialized with 10 randomly selected 
structures, on which the initial cHDNNP was trained.
The same 10 randomly selected structures were used to initialize an RS$_{\mathrm{DFT}}$ training scheme, so that the two schemes were based on the same initial data.
We carried out 10 independent training procedures,
i.e., training 10 cHDNNP with eight committee members each,
for both RS$_{\mathrm{DFT}}$ and AL$_{\mathrm{DFT}}$, each initialized with a different set of randomly selected 
structures,
and report on the average performance of the 10 cHDNNPs in each scheme.

The energy and force root mean square errors (RMSEs) we obtain for the test set containing 667 holdout structures are comparable for the RS$_{\mathrm{DFT}}$ and AL$_{\mathrm{DFT}}$ schemes, at $\sim$0.3~meV/atom for energies, and $\sim$35~meV/{\AA} for forces (Fig. \ref{correlation_rmse}(a, b)).
The usefulness of test RMSEs is limited, because they strongly depend on the
similarity between the training and test sets.
This behavior is related to the fact that the
energy
RMSE is not invariant to a constant shift in the potential energy surface, contrary to, e.g., observables from dynamical simulations.
Therefore, if
an extended training set results in a slightly different optimal offset than random sampling, it will  result in increased test errors due to this bias.
One possible remedy is to use shift-invariant error measures, like force RMSE or Pearson correlation coefficients between predicted and reference energies or forces.
Yet, it is surprising, and perhaps 
contrary to the current understanding of active learning for MLPs,
to find that the active learning scheme produces slightly larger errors than the 
rather basic random sampling 
scheme, even if shift invariant measures are applied (Fig.~\ref{correlation_rmse}(c, d)).

\begin{figure}
\includegraphics{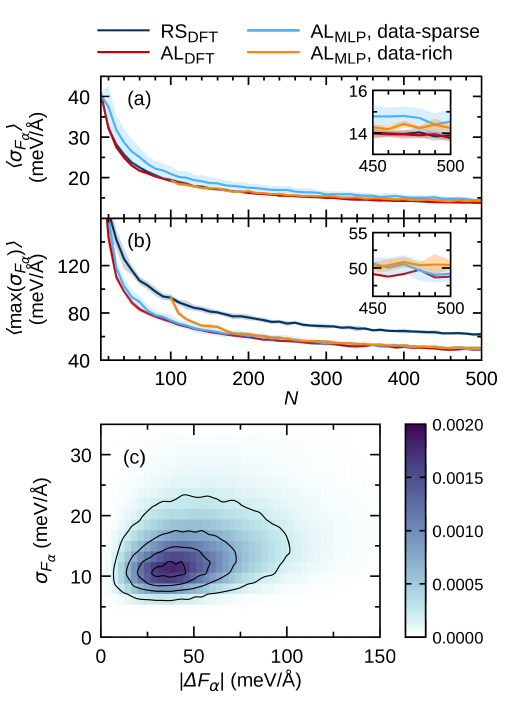}
\caption{
Committee disagreement evaluated on the test set.
(a) Average force committee disagreement in the test set versus number of structures $N$ in the training set, (b) maximum force committee disagreement among atoms in each 128-H$_2$O structure in the test set, averaged over all 667 structures, versus $N$, and (c) the correlation of force committee disagreement of atom $\alpha$, $\sigma_{F_{\alpha}}$, and the error of the cHDNNP force relative to DFT, $| \Delta F_{\alpha} | = | \vec{F}_{\alpha}^{\mathrm{DFT}} - \langle \vec{F}_{\alpha}\rangle |$. The distribution in (c) is normalized to unity, and contour lines are drawn at intervals of 0.0005 starting at 0.0002.
}
\label{correlation}
\end{figure}

We used the maximum force committee disagreement to select new structures during active learning, and indeed $\max_{\alpha}\left(\sigma_{F_{\alpha}}\right)$ decreases more rapidly for the AL$_{\mathrm{DFT}}$ scheme than the RS$_{\mathrm{DFT}}$ with increasing training set size (Fig. \ref{correlation}(b)).
The average force committee disagreement in the test set (Fig. \ref{correlation}(a)) is however similar for AL$_{\mathrm{DFT}}$ and RS$_{\mathrm{DFT}}$.
In using $\max_{\alpha}\left(\sigma_{F_{\alpha}}\right)$ to select new structures during active learning, we have implicitly assumed that those atoms that have large $\sigma_{F_{\alpha}}$ also have a large force error, relative to the DFT ground-truth,
and that including these structures into the training set therefore reduces the discrepancy between the cHDNNP force and the DFT force.
The true error of the cHDNNP force is $| \Delta F_{\alpha} | = | \vec{F}_{\alpha}^{\mathrm{DFT}} - \langle \vec{F}_{\alpha} \rangle |$, where $\vec{F}_{\alpha}^{\mathrm{DFT}}$ is the force on atom $\alpha$ from DFT, and $\langle \vec{F}_{\alpha} \rangle$ is defined in Eq. (\ref{comforce}).
$\left| \Delta F_{\alpha} \right|$ shows a rather moderate correlation with $\sigma_{F_{\alpha}}$ (Fig.~\ref{correlation}(c)), with Pearson correlation coefficient equal to 0.31.
Previous work has also shown that committee disagreement and actual MLP error are not very strongly correlated~\cite{Kahle2022Quality, Zhu2023Fast, Tan2023Single, Zaverkin2024Uncertainty}.
Nevertheless, the correlation is positive, but we have shown that reducing the maximum force committee disagreement more efficiently in AL$_{\mathrm{DFT}}$ versus RS$_{\mathrm{DFT}}$ does not lead to a corresponding reduction of test RMSEs in the AL$_{\mathrm{DFT}}$ schemes.

\begin{figure}[b]
\includegraphics{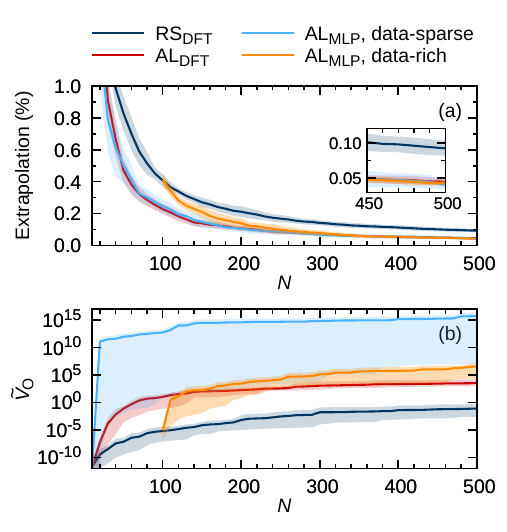}
\caption{Structural diversity in the cHDNNP training sets in each learning scheme, illustrated by (a) the extrapolation instances, in \% 
of all atomic environments 
in the test set, and (b) the volume $\tilde{V}_{\mathrm{O}}$ in high-dimensional symmetry function space included by the symmetry functions describing the oxygen atom environments in the training set, as defined in Eq.~(\ref{hypervolume}), as a function of the number of structures $N$ in the training set.
Solid lines show the average number of extrapolation instances of cHDNNPs (a) and the average of $\tilde{V}_{\mathrm{O}}$ of the training sets (b) from the independent training procedures performed in each of the four training schemes.
The shaded area in (a) is the standard deviation across those cHDNNPs, and the shaded area in (b) indicates minimum and maximum values of $\tilde{V}_{\mathrm{O}}$ in the ensemble of data sets produced in 
each scheme. 
}
\label{extrap}
\end{figure}

\begin{figure*}
\includegraphics[width=0.85\textwidth]{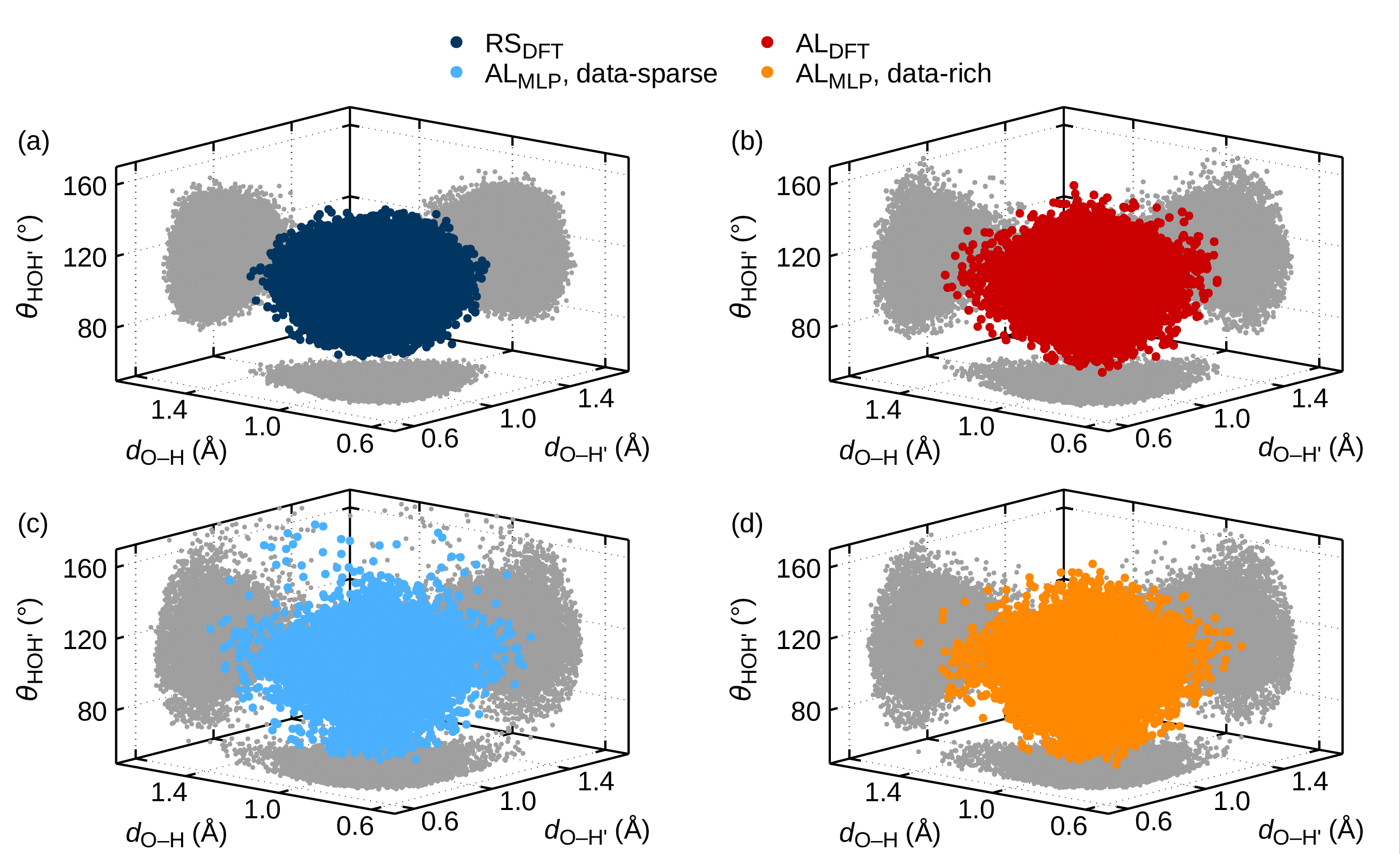}
\caption{
Intramolecular H$_2$O configurations in the training sets constructed according to the different schemes, RS$_\mathrm{DFT}$ (a), AL$_{\mathrm{DFT}}$ (b), data-sparse AL$_{\mathrm{MLP}}$ (c), and data-rich AL$_{\mathrm{MLP}}$ (d).
Points are placed in the scatter plot according to the two covalent bond lengths ($d_{\mathrm{O-H}}$ and $d_{\mathrm{O-H'}}$) and the bond angle $\theta_{\mathrm{HOH'}}$ in each molecule, and the projections of the points on each of the three planes are shown in gray. 
Note that the scales of the axes are increased here with respect to those in Fig. \ref{configurations}(b) as we sampled more configurations.
}
\label{dataset_configurations}
\end{figure*}

\begin{figure}
\includegraphics{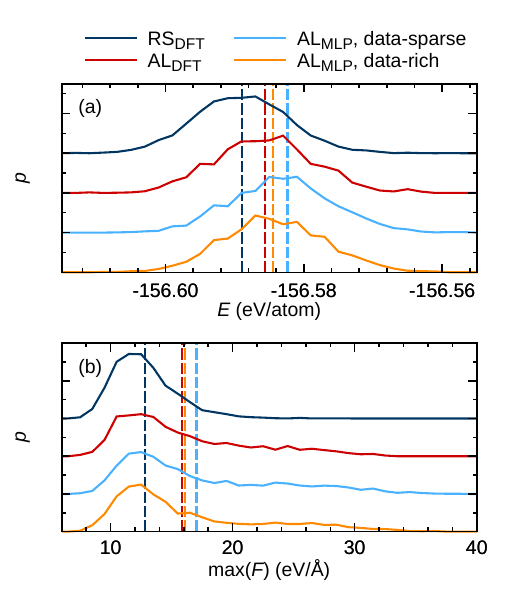}
\caption{Probability distributions
of energies (a) 
and maximum force magnitude in each 128-molecule structure (b) of the training sets constructed in the different schemes. 
The final training sets, containing 500 structures each, for all training procedures performed according to each scheme were pooled together and analyzed.
Dashed vertical lines indicate the mean values of the distributions.
The distributions are each normalized to unity, and for clarity, different distributions are vertically shifted.}
\label{dataset}
\end{figure}

To understand the difference in performance of cHDNNPs trained in AL$_{\mathrm{DFT}}$ and RS$_{\mathrm{DFT}}$ schemes,
which is apparently related to the structures included in the respective training sets,
we quantify the structural diversity of the training sets in two different ways.
Firstly, we look at the number of extrapolation instances encountered by cHDNNPs in the test set (Fig.~\ref{extrap}(a)).
A local atomic environment 
is said to be an extrapolation point when the value of at least one symmetry function lies outside the range of values of that symmetry function in the training set of that cHDNNP \cite{Behler2011Atom, Tokita2023Tutorial}. 
Secondly, we consider the included volume in high-dimensional symmetry function space of the 
atomic environments of each element in the training structures
(Fig.~\ref{extrap}(b)). 
This volume is defined as
\begin{equation} \tilde{V}_{\beta} = \prod_{i=1}^{N_{G^{\beta}}} \left[ \max\left(G^{\beta}_i\right) - \min\left(G^{\beta}_i\right) \right], \label{hypervolume} \end{equation}
where $\beta$ is the element index (O or H), $N_{G^{\beta}}$ is the total number of symmetry functions used for that element, and $G^{\beta}_i$ is the value of the $i$th symmetry function.
The max and min values are taken over all atoms of type $\beta$ in the training set.
It is evident that the
training sets constructed in AL$_{\mathrm{DFT}}$ schemes cover a larger configuration space than those constructed with RS$_{\mathrm{DFT}}$, and consequently  
there are fewer extrapolation instances in AL$_{\mathrm{DFT}}$, and $\tilde{V}_{\beta}$ is larger for both $\beta =$ O and H (the latter is not shown).
The structures in the test set used in Fig.~\ref{correlation_rmse} were drawn from equilibrium PI-AIMD simulations, and as such contain mostly near-equilibrium structures, which are well-represented in the RS$_{\mathrm{DFT}}$ training set (Fig.~\ref{dataset_configurations}(a)), while the AL$_{\mathrm{DFT}}$ training set contains \emph{both} near-equilibrium and more distorted local water structures (Fig.~\ref{dataset_configurations}(b)), 
for which the committee disagreement tends to be larger. 
The larger structural diversity in the active learning training sets is also reflected in the
larger range of 
training energies and force magnitudes (Fig.~\ref{dataset}).
Therefore, in the AL$_{\mathrm{DFT}}$ scheme, the cHDNNP has to fit a larger range of structures and their
associated energies and forces, and that slightly deteriorates the quality of the fit in near-equilibrium regions of the potential.
Still, the inclusion of extremely distorted structures is avoided in this active learning scheme, as structures are selected from the pool of PI-AIMD data that contains overall reasonable atomic configurations.

Incorporating more distorted structures in the training set may be assumed to be necessary to achieve an MLP that is long-term stable in simulations;
that way, repulsive regions are sampled well and simulations will remain close to physically relevant regions.
However, we will see 
below that beyond very small training sets containing fewer than 100 periodic 
structures of 128 water molecules, the quality of simulations is independent of the structural composition of the training sets for bulk liquid water.

\subsection*{Active learning from MLP structures}

While the RS$_{\mathrm{DFT}}$ and AL$_{\mathrm{DFT}}$ procedures outlined above are suitable for the investigation of the relevance of high-uncertainty structures in MLPs, and active learning procedures similar to the one investigated here have been used in the past \cite{Sivaraman2020Machine, Schran2021Machine}, 
these selection procedures are based on structures drawn from an existing \textit{ab initio} trajectory.
In the case where such a long reference trajectory is not available, 
for example because it is impossible to sample the full configuration space with \textit{ab initio} methods,
new structures could be sampled from simulations with an intermediate MLP. 
Many examples of such procedures exist, including for water~\cite{Raff2005AbIntio, Malshe2007Theoretical, Artrith2012High-dimensional, Morawietz2012Neural, Sosso2012Neural, Morawietz2013Density-Functional, Morawietz2016How, Deringer2017Machine, Schran2020Automated, Schran2020Committee, Daru2022Coupled, Kapil2022First-Principles, Tang2023Machine, ONeill2024Pair}.
In these procedures, if the intermediate potential is of high quality, then new structures that are added to the training set are already close to structures from \textit{ab initio} simulations, and thus we expect to see little difference between the performance of a potential trained on \textit{ab initio} or MLP structures. 
If, however, the intermediate potential is of lower quality, new suggested structures may deviate significantly from those in the target distribution, and may even be unphysical. 
In that case, the training set will contain structures of little relevance to the target system, and it is unclear whether the MLP in subsequent training iterations will attain a good quality, or even a physically sound description, of the system of interest.
Thus, here for comparison we also explore such learning schemes, termed AL$_{\mathrm{MLP}}$ (schemes 3 and 4 in Fig.~\ref{schemes}).

We started with an underconverged cHDNNP 
trained only to a few initial randomly chosen structures,
and ran 5 independent 80~ps path integral simulations with the cHDNNP.
The first 20~ps of each trajectory were discarded, leaving a total trajectory length of \mbox{5 $\times$ 60 ps = 300 ps}.
From those trajectories, frames were drawn at 30~fs intervals, and the force committee disagreement (Eq. (\ref{comdis})) was evaluated for a randomly selected path integral bead from each frame.
The 10 structures with the largest $\max_{\alpha}\left(\sigma_{F_{\alpha}}\right)$ were then added to the cHDNNP training set.
The cHDNNP was re-trained, and new path integral simulations were performed to select structures for the next step
of the training procedure. 

Here, we compare the convergence and performance of cHDNNPs trained using active learning seeded with a cHDNNP initially trained on 10 
structures
which were selected at random from the PI-AIMD trajectory, and cHDNNPs trained in procedures seeded with a cHDNNP trained on 100 randomly selected 
structures.
The former sets of cHDNNPs are referred to as ``AL$_{\mathrm{MLP}}$, data-sparse", and the latter ``AL$_{\mathrm{MLP}}$, data-rich".
For each of these cases, we performed 5 independent training procedures, each initialized with a different set of randomly selected structures.
We note that this procedure, i.e., sampling structures from entire preliminary cHDNNP-generated trajectories, is a challenging setup as in case of lower quality initial potentials the system might be driven away from the physically relevant configuration space in the course of the simulation.
To avoid this issue, often only uncertain structures visited early in the trajectories, which are thus not too different from the starting structures, are selected \cite{Behler2015Constructing}.
Here, on purpose, instead we follow a fully automated procedure and sample structures from the entire trajectory without judging on the individual atomic configurations generated before including them in the training set, and the only criterion is the committee disagreement.

\begin{figure}
\includegraphics{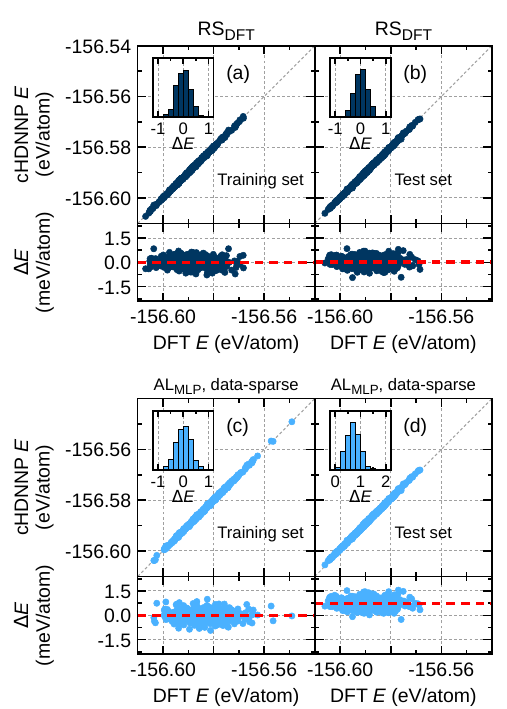}
\caption{Correlation plots of cHDNNP energy predictions and DFT energies, for the RS$_{\mathrm{DFT}}$ scheme (a, b) and the AL$_{\mathrm{MLP}}$ data-sparse scheme (c, d). 
The lower panels show energy differences $\Delta E$ between DFT and cHDNNP energies, and insets show histograms of the energy differences.
Dashed horizontal red lines indicate the average of $\Delta E$ for each case.
Energies are computed for the final training sets containing 500 128-H$_2$O structures (a, c), and the test set containing 667 128-H$_2$O structures from the PI-AIMD trajectory (b, d).
The training sets are different for each training procedure, but the test set is the same.
}
\label{e_correlations}
\end{figure}

\begin{figure*}
\includegraphics{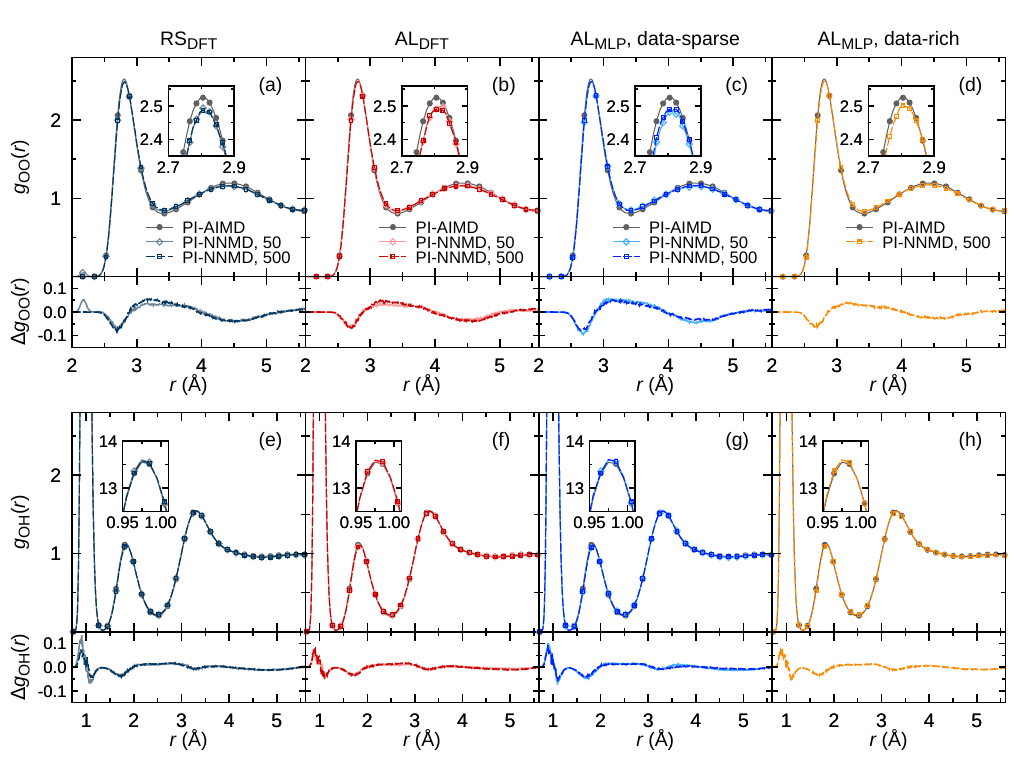}
\caption{Partial OO (a--d) and OH (e--h) RDFs from PI-NNMD simulations using
cHDNNPs trained based on RS$_{\mathrm{DFT}}$ (a, e), AL$_{\mathrm{DFT}}$ (b, f), and AL$_{\mathrm{MLP}}$ with data-sparse (c, g) and data-rich (d, h) starts.
The numbers in the legends indicate the number of 128-H$_2$O structures in the training set of the cHDNNP used in simulations. For reference, also PI-AIMD RDFs are shown, and the $\Delta g$ subpanels show the difference between the PI-NNMD RDFs and the relevant PI-AIMD RDFs.
The radial bin size is 0.01 {\AA} and distributions are not smoothened.
For clarity, a subset of data points are represented with symbols in the RDFs.
}
\label{rdfs}
\end{figure*}

Structures contained in AL$_{\mathrm{MLP}}$ training sets are significantly more diverse than those in random and active learning training sets (Fig. \ref{extrap}), and also tend to have higher energies and larger forces (Fig. \ref{dataset}).
Furthermore, the AL$_{\mathrm{MLP}}$ scheme with data-sparse start led to considerably more structural diversity in the training set than the data-rich start, because the initial cHDNNPs were very poorly fitted to bulk liquid water and therefore many distorted configurations were drawn from the PI-NNMD simulations for addition to the training set (Fig.~\ref{dataset_configurations}(c, d)).
In fact, for 1 out of the 5 data-sparse AL$_{\mathrm{MLP}}$ training procedures, new selected training structures were so distorted that they no longer resembled liquid water, and the \mbox{RPBE-D3} calculations to obtain the DFT energies and forces did not converge.
Therefore, we stopped this particular training procedure, and results reported for the data-sparse AL$_{\mathrm{MLP}}$ scheme refer to an ensemble of the 4 remaining training procedures only.

Due to the increased complexity of the underlying data sets,
both AL$_{\mathrm{MLP}}$ learning schemes produce significantly larger energy RMSEs than the AL$_{\mathrm{DFT}}$ and RS$_{\mathrm{DFT}}$ schemes, while the errors in forces are comparable (Fig.~\ref{correlation_rmse}(a,b)).
In fact, as is particularly clear from the data-rich AL$_{\mathrm{MLP}}$ procedure, the energy RMSE increases once PI-NNMD structures are added to the training set.
Since the intermediate cHDNNPs are not fully converged and therefore do not exactly reproduce the PI-AIMD ensemble, the data points that were added to the training set in the AL$_{\mathrm{MLP}}$ schemes have, on average, a higher energy than those added in our AL$_{\mathrm{DFT}}$ and RS$_{\mathrm{DFT}}$ schemes, where data points were selected from the 
equilibrated PI-AIMD trajectory (Fig.~\ref{dataset}).
The AL$_{\mathrm{MLP}}$ cHDNNPs do fit their training data correctly, with similar quality as achieved in the RS$_{\mathrm{DFT}}$ scheme (Fig.~\ref{e_correlations}); 
however, these training data are structurally subtly different from the structures in the test set.
Ultimately, this has led to an offset in the energies predicted by the cHDNNPs trained on these structures, relative to RPBE-D3 energies, for the test set made up of frames drawn from the PI-AIMD trajectory.
However, the correlation coefficient between RPBE-D3 and cHDNNP energies, which is insensitive to a constant offset, is close to those from AL$_{\mathrm{DFT}}$ and RS$_{\mathrm{DFT}}$ schemes, and increases more or less monotonically with training set size (Fig.~\ref{correlation_rmse}(c)).
Additionally, the force RMSE decreases with increasing training set size, as gradients are unaffected by the offset in energies.
We conclude that the larger structural diversity and hence higher average energy in AL$_{\mathrm{MLP}}$ training sets leads to an offset in cHDNNP energy predictions for the particular test set that is used here,
but that this does not affect forces and thus the performance of the cHDNNPs in molecular dynamics simulations.
Energy RMSEs are often used as a signifier of MLP quality, but it may be more appropriate to consider correlation coefficients in applications where energy differences rather than absolute energy values are of relevance.

\subsection*{End-to-end testing}

While test RMSEs are commonly used to assess MLPs, these may not be representative of their performance~\cite{Morrow2023How}. 
Instead, in order to understand the quality of MLPs, they should be used in tests that are physically guided and specific to the final application of the MLPs.
In this work, we have aimed to train cHDNNPs that reproduce the structure of quantum liquid water at the RPBE-D3 level of theory.
Therefore, we performed 420 ps long constant-volume, constant-temperature PI-NNMD simulations with intermediate and final cHDNNPs produced in all learning schemes, and compare the radial structure to that from the PI-AIMD trajectory. 
The first 20 ps were discarded as equilibration.
We computed RDFs from these simulations to assess the quality of the cHDNNPs (Fig.~\ref{rdfs}).
With only 50 structures in the training sets, and in fact even in most cases with only 10 structures in the training set, the cHDNNPs produce stable simulations that already have a reasonable water structure, as probed by the RDFs.
Note that we do not show results with 50 structures for the data-rich AL$_{\mathrm{MLP}}$ scheme, since that scheme was initialized with 100 structures.
In particular the intramolecular O--H peak is already well-described with the smaller training sets in all cases, since all training structures contain detailed information about the intramolecular structure of H$_2$O including thermal fluctuations.
With 500 structures in the training sets, 
all RDFs are in excellent agreement with the PI-AIMD reference.

The only major difference between the different learning schemes is that with a small training set with 50 128-H$_2$O structures, the RS$_{\mathrm{DFT}}$ scheme produces a small peak in the O--O RDF at short distances where the intermolecular interaction should be repulsive (Fig. \ref{rdfs}(a)), 
which is not sufficiently learned from very small random data sets.
These close contacts of water molecules are avoided with the smaller training sets produced with active learning.
Selection of new structures based on committee disagreement aids cHDNNPs in sampling those repulsive regions when the training set is particularly small.
Still, the  quality of the RDFs produced with cHDNNPs trained on 500 structures is exceedingly similar across learning schemes, indicating that eventually, the RS$_{\mathrm{DFT}}$ scheme also constrains the repulsive regions appropriately.

\begin{figure}
\includegraphics{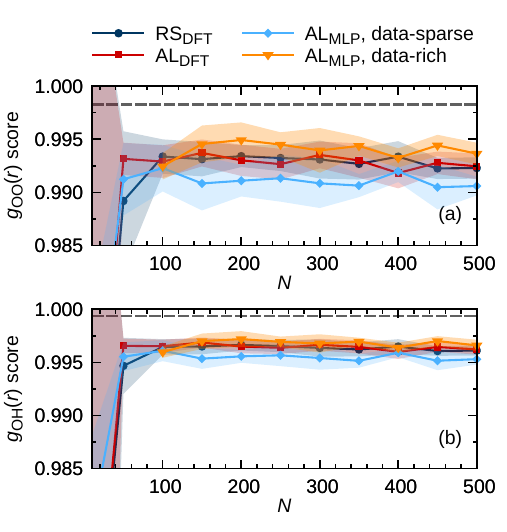}
\caption{Similarity scores of the cHDNNP RDFs $g_{\mathrm{OO}}$ (a) and $g_{\mathrm{OH}}$ (b) from PI-NNMD simulations relative to the PI-AIMD RDFs, computed according to Eq.~(\ref{score_eq}), as a function of the number of structures $N$ in the training set. 
Solid lines show average similarity scores of 
the
cHDNNPs from independent training procedures performed in each of the four learning schemes,
and the shaded areas are the standard deviations of those scores.
The dashed horizontal lines 
indicate the similarity score of RDFs computed from the first and second halves of the PI-AIMD trajectory, which is thus the expected similarity score for RDFs produced by exactly the same method but with finite statistical sampling.
}
\label{scores}
\end{figure}

To quantify the convergence of the cHDNNP quality in path integral simulations, we use a similarity score between the reference PI-AIMD RDFs and the PI-NNMD RDFs,
\begin{equation}
\mathrm{score} = 1 - \frac{\int_0^{L/2} \left| g^{\text{PI-AIMD}}(r) - g^{\text{PI-NNMD}}(r) \right| dr}{\int_0^{L/2} g^{\text{PI-AIMD}}(r)dr + \int_0^{L/2} g^{\text{PI-NNMD}}(r)dr},
\label{score_eq}
\end{equation}
where $L/2$ is half the simulation cell size. 
This score is equal to 1 when the PI-AIMD and PI-NNMD RDFs are exactly equal, and smaller otherwise.
For our learning schemes, the RDF scores show a large variation even within a learning scheme initially; cHDNNPs trained on small training sets are in general poorly constrained and may or may not produce sensible water structures (Fig.~\ref{scores}).
Beyond approximately 200 
structures in the training sets, the scores converge, indicating that adding more structures does not have any 
significant effect on the performance of cHDNNPs in simulations.

The RDF scores also reveal that the data-sparse AL$_{\mathrm{MLP}}$ scheme on average produces cHDNNPs that perform slightly worse in simulations than the other learning schemes.
This conclusion could not be readily drawn from, e.g., the RMSEs (Fig.~\ref{correlation_rmse}), and the larger structural diversity in the training sets produced with the data-sparse AL$_{\mathrm{MLP}}$ scheme (Fig.~\ref{extrap}) could even have hinted at 
a potentially better performance of these cHDNNPs.
From the RDF scores, we see that the distorted and at times unphysical structures added to the cHDNNP training sets are detrimental to the performance of these cHDNNPs.
We also recall that 1 
cHDNNP committee out of 5 attempted trained with the data-sparse scheme failed to produce sensible water structures in simulations with the initial cHDNNPs, so that the procedure was stopped altogether.
Therefore, while AL$_{\mathrm{MLP}}$ is a strategy to avoid long (PI-)AIMD simulations when training MLPs, it does rely on a reasonable starting MLP,
requiring a certain minimum number of physically meaningful structures.
Our data-rich AL$_{\mathrm{MLP}}$ learning scheme, initialized with 100 frames rather than 10, performed well for that reason.
Here, initial cHDNNPs were trained on structures from an \textit{ab initio} simulation,
but we speculate that one might use reasonable structures produced with a lower level of theory, e.g., a 
good-quality flexible force field.

At this point, it is worth noting that 
in the literature a lot of importance is typically attached to energy and force RMSEs of MLPs, to assess their quality, 
in particular when new MLP methods are introduced.
Yet, we show here that cHDNNPs that produce energy RMSEs that are more than twice that of other cHDNNPs still perform 
equally well in simulations, remaining close to the target level of theory.
The reported RMSEs are inherently sensitive to the composition of the test set, 
which in any case is very small compared to the huge number of atomic configurations encountered in long MD trajectories;
as a consequence of the particular procedure used to construct the training set on which the cHDNNP is trained, the training set may contain structures, and energies, which are not entirely representative of the structures in the test set used to assess RMSEs, without deteriorating the quality of the cHDNNP in simulations.
Furthermore, the energy gradients are insensitive to absolute energy errors, and models can provide the right physics in molecular dynamics simulations, which rely on forces, 
in spite of somewhat larger energy RMSEs.
Here, we have therefore 
in addition reported energy and force correlation coefficients of the test set beside RMSEs (Fig.~\ref{correlation_rmse}(c,d)), but the true test of an MLP remains its performance for the target application---here, path integral molecular dynamics simulations,
and more specifically the water structure from those simulations, as probed by the RDFs.

We have shown that active learning does not improve the performance of cHDNNPs in PI-NNMD simulations of bulk liquid water beyond what random sampling can achieve.
To arrive at this conclusion, we have performed 20 separate training procedures across four different learning schemes, while assessing intermediate and final cHDNNPs at the level of PI-NNMD simulations, until the training sets contained 500 structures.
Still, our active learning schemes are based on maximum force committee disagreement, but there are many other ways in which new structures could be selected for addition to the training set.
For example, we might rely on relative uncertainties rather than absolute uncertainties to select new data, look for extrapolation instances to flag relevant training structures, use real-space information of candidate structures in e.g. farthest point sampling, or use a combination of active learning and random sampling.
Beyond the structure selection criterion, there are countless degrees of freedom in the training set construction algorithm, and it is unattainable to carry out the careful analysis that we have here for all possible cases.
Nevertheless, our RS$_{\mathrm{DFT}}$ and AL$_{\mathrm{DFT}}$ schemes provide a balanced comparison between active learning and random sampling, while our AL$_{\mathrm{MLP}}$ schemes investigate a perhaps more common application of active learning.

\section*{Conclusions and Outlook}

By comparing four different schemes for construction of training sets for cHDNNPs for quantum liquid water, we have shown that ``smart" selection of new training structures does not lead to cHDNNPs of better quality than random selection.
Instead, random
sampling 
and the various active learning schemes all lead to high-quality cHDNNPs with surprisingly moderate training set sizes, with the caveat that AL$_{\mathrm{MLP}}$ must be initialized with a reasonable model.
We compared the quality of the fitted networks based on energies, forces and structural properties.
According to 
our analysis, energy-based error measures have to be invariant with respect to constant shifts in order to provide consistent results with the quality of the structural properties.
These findings complement existing work that has found good-quality MLPs for liquid water and aqueous systems using various active learning schemes.
Recent MLP methods, e.g., employing equivariant descriptors and message passing, have been demonstrated 
to 
work with small data sets~\cite{Batzner2022E3, Kovacs2023Evaluation},
but indeed for quantum liquid water, second-generation HDNNPs achieve excellent quality with $\sim$200 structures in the training set as well.

Training an MLP by randomly selecting structures from an existing 
\textit{ab initio} data base is significantly more efficient than the AL$_{\mathrm{DFT}}$ scheme that we have used, because it does not require re-training of the MLP after adding a new batch of structures.
Instead, single-shot training is sufficient and random learning should thus be preferred 
for those systems for which \textit{ab initio} trajectories covering the entire configuration space of interest are available.

When no existing data base of structures is available, AL$_{\mathrm{MLP}}$ schemes 
offer an alternative option to sample the relevant configuration space.
In that case, new structures must be generated at every step as well, for example with molecular dynamics simulations. 
Our AL$_{\mathrm{MLP}}$ could be made more computationally efficient, e.g., by adding more structures at each sampling step, by running shorter simulations, or by adding new structures as soon as MLP predictions become uncertain or are extrapolating \cite{Behler2015Constructing}.

Our analyses in this work have been performed for bulk liquid water at ambient conditions.
As outlined in the introduction, quantum liquid water is a relatively challenging system, but it benefits from 
configurational
self-similarity.
A single frame from a simulation of liquid water contains many water molecules in their local environment, information that can be used to fit the parameters of an MLP.
For that reason, not many frames are needed for the covalent O--H bond to be correctly described by our cHDNNPs (Fig. \ref{rdfs}(e--h)).
We expect that random learning leads to equivalent results as active learning for isotropic molecular liquids or solids.
On the other hand, extended molecules, systems that can undergo large-amplitude rearrangements or phase transitions, or anisotropic systems such as interfaces, may require more sophisticated approaches to training set construction
and are probably less suited for random sampling from AIMD trajectories as these can be restricted to sampling structures in specific basins of the potential energy surface.

In this work, we have not attempted to find the most sophisticated (most data efficient, easiest to implement, fastest, etc.) active learning procedure for the construction of training sets for MLPs.
Instead, we have focused on one particular machine learning method for atomistic simulations, namely cHDNNPs, and transparently compared active learning and random sampling.
The final quality of our cHDNNPs is limited not by the training sets, as illustrated by the convergence of the various quality criteria that we have investigated (Fig. \ref{correlation_rmse}--\ref{extrap}, \ref{scores}), but may be limited by the flexibility of the cHDNNPs due to
specific choice of architecture and symmetry functions, and by possible noise in the training set due to finite convergence of the electronic structure calculations. 
Still, in spite of not systematically varying these parameters, high-quality potentials have been obtained.
The exact convergence behavior may look qualitatively different with different machine learning methods or data sources, 
and of course also for different systems.
Yet, we show that for a popular choice of MLP (HDNNP) and uncertainty quantification (query by committee), active learning does not outperform random learning.
The training data selected in active learning algorithms covers a larger configurational space than the training data selected using random sampling, but that does not improve the performance of cHDNNPs from AL$_{\mathrm{DFT}}$ or AL$_{\mathrm{MLP}}$ in simulations.
In fact, the higher-energy configurations in active learning training sets led to larger test RMSEs than produced by random sampling.
We anticipate that our findings stimulate discussion around the impact of existing active learning strategies in training MLPs, and further encourage the development of proven methods to reduce training set sizes.

\section{Acknowledgments}
Funded by the Deutsche Forschungsgemeinschaft (DFG, German Research Foundation) under Germany's Excellence Strategy~-- EXC~2033~-- 390677874~-- RESOLV.
Financial support from the National Research, Development and Innovation Office (NKFIH, Grant No. FK147031) is gratefully acknowledged by JD. JB is grateful for support by the DFG (BE3264/16-1, project number 495842446 in priority program SPP 2363 ``Utilization and Development of Machine Learning for Molecular Applications -- Molecular Machine Learning'').
All computations have been carried out locally at HPC@ZEMOS, HPC-RESOLV, and BOVILAB@RUB. 

\bibliography{main}

\end{document}


\title{Supporting information for:\\Random sampling versus active learning algorithms for machine learning potentials of quantum liquid water}

\author{Nore Stolte}
\affiliation{Lehrstuhl f{\"u}r Theoretische Chemie, Ruhr-Universit{\"a}t Bochum, 44780 Bochum, Germany}
\author{J{\'a}nos Daru}
\affiliation{Lehrstuhl f{\"u}r Theoretische Chemie, Ruhr-Universit{\"a}t Bochum, 44780 Bochum, Germany}
\affiliation{Department of Organic Chemistry, E{\"o}tv{\"o}s Lor{\'a}nd University, 1117 Budapest, Hungary}
\author{Harald Forbert}
\affiliation{Center for Solvation Science ZEMOS, Ruhr-Universit{\"a}t Bochum, 44780 Bochum, Germany}
\author{Dominik Marx}
\affiliation{Lehrstuhl f{\"u}r Theoretische Chemie, Ruhr-Universit{\"a}t Bochum, 44780 Bochum, Germany}
\author{J{\"o}rg Behler}
\affiliation{Lehrstuhl f{\"u}r Theoretische Chemie II, Ruhr-Universit{\"a}t Bochum, 44780 Bochum, Germany}
\affiliation{Research Center Chemical Sciences and Sustainability, Research Alliance Ruhr, 44780 Bochum, Germany}

\maketitle

\section{Symmetry functions}

We use radial symmetry functions of the form 
\begin{equation}
G_i^{\mathrm{rad}} = \sum_j e^{- \eta (r_{ij}-r_s)^2} \cdot f_c(r_{ij}) ,\end{equation}
where $i$ labels the central atom and the sum is over all neighboring atoms.
$r_{ij}$ is the distance between atom $i$ and $j$, and
$f_c(r_{ij})$ is a function which varies smoothly to 0 at the cutoff,
\begin{align} 
f_c(r) = \left\{ \begin{aligned}
& \; \frac{1}{2} \left[ \cos \left( \frac{\pi r}{r_c}\right) + 1\right] & \text{for\;} r \leq r_c \;,\\
& \; 0 & \text{otherwise} .
\end{aligned}
\right.
\end{align}
The angular symmetry functions have form
\begin{equation}
G_i^{\mathrm{ang}} = 2^{-\zeta} \sum_j \sum_{k \neq j} \left[ 1 + \lambda \cos \left(\theta_{ijk} \right) \right]^{\zeta} \cdot e^{\eta \left( r_{ij}^2 + r_{ik}^2 + r_{jk}^2 \right)} f_c(r_{ij})f_c(r_{ik})f_c(r_{jk}) ,
\end{equation}
where the sums are over all neighboring atoms.
Table \ref{radial} and \ref{angular} show the sets of values chosen for $\eta$, $r_s$, $\lambda$, and $\zeta$ in the radial and angular functions for O and H atoms.
$r_c$ is set to 12~$a_0$ in all cases.

\begin{table}[h]
\centering
\caption{Radial symmetry function parameters. Each value for $\eta$ and $r_s$ in the set in braces combines individually with the value for $r_s$ or $\eta$ on the same row to give a radial function.}
\label{radial}
\begin{tabular}{c c c} 
\hline
\hline
\; Element pair \; & $\eta$ ($a_0^{-2}$) & $r_s$ ($a_0$) \\
\hline 
\hline
H--O & $\left\lbrace 0.0,0.01,0.03,0.06,0.15 \right\rbrace$ & 0.0 \\ 
 & 2.0 & $\left\lbrace 0.0,1.5,3.0,4.5,6.0,7.5,9.0 \right\rbrace$ \\ 
\hline
O--O & $\left\lbrace 0.0,0.01,0.03,0.06,0.15 \right\rbrace$ & 0.0 \\ 
 & 2.0 & $\left\lbrace 4.5,6.0,7.5,9.0 \right\rbrace$ \\ 
\hline
H--H & $\left\lbrace 0.0,0.01,0.03,0.06,0.15 \right\rbrace$ & 0.0 \\ 
 & 2.0 & $\left\lbrace 1.5,3.0,4.5,6.0,7.5,9.0 \right\rbrace$ \\ 
\hline
\hline
\end{tabular}%
\end{table}

\begin{table}[t]
\centering
\caption{Angular symmetry function parameters. Each value for $\eta$, $\lambda$ and $\zeta$ in the sets in curly braces combines in turn with each other value in the sets on each row to give the angular symmetry functions.}
\label{angular}
\begin{tabular}{c c c c c} 
\hline
\hline
\; Central atom \; & \; Neighbor atoms \; & $\eta$ ($a_0^{-2}$) & $\lambda$ & $\zeta$ \\
\hline 
\hline
H & H, O & $\left\lbrace 0.0, 0.030, 0.100\right\rbrace$ & $\left\lbrace -1,1\right\rbrace$ & $\left\lbrace 1,2,4,16 \right\rbrace$ \\
\hline
H & O, O & $\left\lbrace 0.0, 0.030, 0.100\right\rbrace$ & $\left\lbrace -1,1\right\rbrace$ & $\left\lbrace 1,2,4,16 \right\rbrace$ \\
\hline
O & H, H & $\left\lbrace 0.0, 0.030, 0.100\right\rbrace$ & $\left\lbrace -1,1\right\rbrace$ & $\left\lbrace 1,2,4,16 \right\rbrace$ \\
\hline
O & H, O & $\left\lbrace 0.0, 0.030, 0.100\right\rbrace$ & $\left\lbrace -1,1\right\rbrace$ & $\left\lbrace 1,2,4,16 \right\rbrace$ \\
\hline
O & O, O & $\left\lbrace 0.0, 0.015, 0.040\right\rbrace$ & $\left\lbrace -1,1\right\rbrace$ & $\left\lbrace 1,2,4,16 \right\rbrace$ \\
\hline
\hline
\end{tabular}%
\end{table}